\documentclass[prx,superscriptaddress,longbibliography,twocolumn]{revtex4-1}
\usepackage{graphicx}
\usepackage{bm}
\usepackage{xfrac}
\usepackage{xcolor}

\newcommand{\be}{\begin{equation}}
\newcommand{\ee}{\end{equation}}
\newcommand{\bk}{{{\bf{k}}}}

\newcommand{\bea}{\begin{eqnarray}}
\newcommand{\eea}{\end{eqnarray}}
\newcommand{\beal}{\begin{align}}
\newcommand{\eeal}{\end{align}}

\begin{document}
%

\title{Dirac magnons in a honeycomb lattice quantum XY  magnet CoTiO$_3$}

\author{Bo Yuan}
\affiliation{Department of Physics, University of Toronto, Toronto, Ontario, M5S 1A7, Canada}

\author{Ilia Khait}
\affiliation{Department of Physics, University of Toronto, Toronto, Ontario, M5S 1A7, Canada}

\author{Guo-Jiun Shu}
\affiliation{Department of Materials and Mineral Resources Engineering, National Taipei University of Technology, Taipei 10608, Taiwan}
\affiliation{Institute of Mineral Resources Engineering, National Taipei University of Technology, Taipei 10608, Taiwan}
\affiliation{Taiwan Consortium of Emergent Crystalline Materials, Ministry of Science and Technology, Taipei 10622, Taiwan}

\author{F. C. Chou}
\affiliation{Center for Condensed Matter Sciences, National Taiwan University, Taipei, 10617 Taiwan}

\author{M. B. Stone}
\affiliation{Neutron Scattering Division, Oak Ridge National Laboratory, Oak Ridge, Tennessee 37831, USA}

\author{J. P. Clancy}
\affiliation{Department of Physics and Astronomy, McMaster University, Hamilton, ON L8S 4M1 Canada}

\author{Arun Paramekanti}
\affiliation{Department of Physics, University of Toronto, Toronto, Ontario, M5S 1A7, Canada}

\author{Young-June Kim}
\affiliation{Department of Physics, University of Toronto, Toronto, Ontario, M5S 1A7, Canada}

\begin{abstract}
The discovery of massless Dirac electrons in graphene and topological Dirac-Weyl materials
has prompted a broad search for bosonic analogues of such Dirac particles.
Recent experiments have found evidence for Dirac magnons above an Ising-like ferromagnetic ground state in a
two-dimensional (2D) kagome lattice magnet and in the van der Waals layered  honeycomb crystal CrI$_3$, and
in a 3D Heisenberg magnet Cu$_3$TeO$_6$.
Here we report on our inelastic neutron scattering investigation on large single crystals of a stacked honeycomb lattice magnet CoTiO$_3$, which is part of a broad family of
ilmenite materials.
The magnetically ordered ground state of CoTiO$_3$ features ferromagnetic layers of Co$^{2+}$, stacked antiferromagnetically along the $c$-axis. We discover
that the magnon dispersion relation exhibits strong easy-plane exchange anisotropy and hosts a clear gapless Dirac cone
along the edge of the 3D Brillouin zone.
Our results establish CoTiO$_3$ as a model pseudospin-$1/2$ material to study interacting Dirac bosons in a 3D quantum XY magnet.
\end{abstract}

\maketitle


The discoveries of graphene and topological insulators have led to significant advances in our understanding of the properties of electron in solids described by a Dirac
equation. In particular, the fruitful analogy between fundamental massless Weyl-Dirac fermions in Nature and electrons in graphene or topological semimetals
has allowed physicists to simulate theories of particle physics using tabletop experiments \cite{Novoselov2005,Zhang2005,Katsnelson2006,graphene_review,Wehling2014,topologicalphase_review}.
Remarkably, the concept of Dirac particles is not limited to electrons or other fermionic quasiparticles, prompting a search for analogues
in photonic crystals \cite{Lu2014,topologicalphoton_review}, acoustic metamaterials \cite{Jin2018}, and quantum magnets \cite{Fransson2016,Owerre2016,Pershoguba2018,Boyko2018}.
In particular, Dirac magnons, or more broadly defined topological magnons \cite{Zhang2013,Mook2014,Chisnell2015,Chen2018,Yao2018,Lu2018},
have attracted much attention
as platforms to investigate the effect of inter-particle interaction or external perturbations on Dirac bosons, and are proposed to be of
potential interest in spintronic applications.

In contrast to light and sound,
the symmetry broken states and emergent bosonic excitations of quantum magnets depend crucially on dimensionality and spin symmetry, which provides a fertile playground
for examining the physics of topological bosons.
To date, gapped topological magnons in Ising-like ferromagnets
have been reported in a kagome lattice material Cu(1,3-bdc)\cite{Chisnell2015} and in a layered honeycomb magnet CrI$_3$\cite{Chen2018}.
On the other hand, magnons exhibiting symmetry protected band crossings have been found only in a single material, a three-dimensional
(3D) Heisenberg antiferromagnet Cu$_3$TeO$_6$
\cite{Yao2018, Bao2018}. It is thus desirable to explore new test-beds with distinct spin symmetries to expand our understanding of
the physics of Dirac magnons.

In this paper, we present a new model 3D quantum XY magnet realizing gapless Dirac magnons, CoTiO$_3$, which has a simple ilmenite crystal structure. The magnetic lattice of Co$^{2+}$ ions in CoTiO$_3$ is a stacked honeycomb lattice, exactly the same as in ABC stacked graphene. Below $T_N \approx 38$\,K, this material exhibits magnetic order
with ferromagnetic planes stacked antiferromagnetically along the ${\bf c}$-axis.
Our inelastic neutron scattering (INS) experiments reveal crystal field excitations and sharp low energy dispersive spin waves.
Although Co$^{2+}$ is a spin $S\!=\! 3/2$ ion, resulting in a large magnetic signal, our analysis of the observed low energy crystal field levels provides evidence for
strong easy-plane single-ion anisotropy leading to a pseudospin $\tilde{S}=\frac{1}{2}$ doublet ground state.
The low energy spin-wave dispersion reveals clear crossings of magnon modes exhibiting Dirac cone dispersion in the 3D Brillouin zone (BZ).
A simple model Hamiltonian with dominant nearest neighbor XY ferromagnetic (FM)
exchange, and antiferromagnetic interlayer second-neighbor exchange, provides an excellent description of
the magnon dispersion.
In addition, the observed spin-wave intensities highlight the stark contrast between the
two different transverse fluctuation modes, distinguishing this XY magnet from Ising-like or Heisenberg magnets.

Unlike the other materials mentioned above, large, pristine, single crystal samples of this material, suitable for neutron scattering and other experiments,
can be grown using the conventional floating zone method.
In addition, ilmenites can be grown as epitaxial thin films using conventional oxide film growth methods \cite{Varga2012}, permitting
future explorations of epitaxial strain and incorporation in spintronic devices.

\begin{figure}[tb]
	\centering
\includegraphics[width=0.45\textwidth]{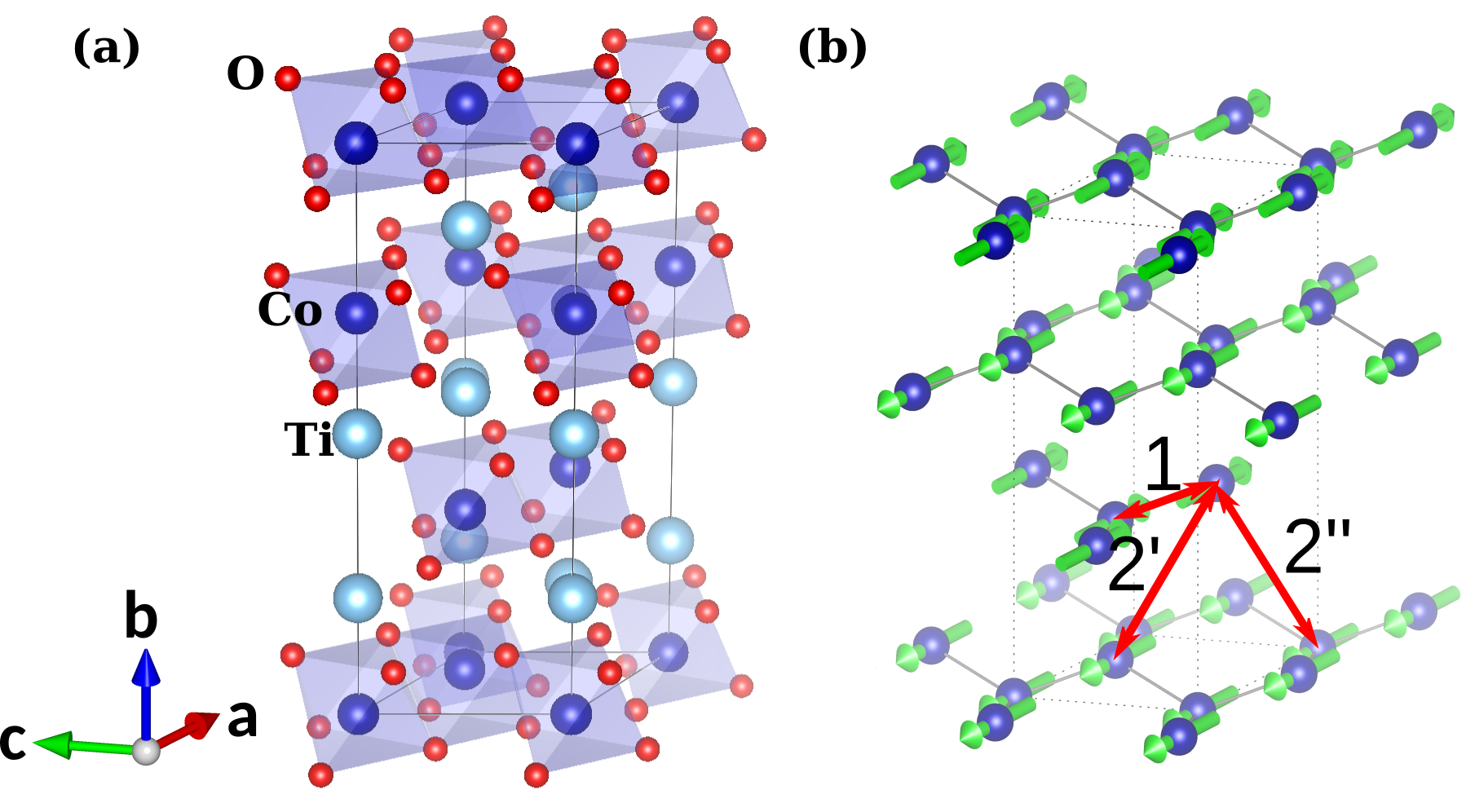}
	\caption{(a) Structure of CoTiO$_3$. Non-primitive hexagonal unit cell used to describe our data is indicated by black solid lines. (b) Magnetic
	structure of the Co$^{2+}$ sublattice. Co$^{2+}$ magnetic moments (shown in green arrows) order ferromagnetically within each honeycomb plane and
	antiferromagnetically along the ${\bf c}$-axis. Red arrows labelled 1, $2^\prime$ and $2^{\prime\prime}$ are the magnetic couplings considered
	in the spin wave calculation.}
	\label{structure}	
\end{figure}

{\bf Crystal structure, crystal field levels:}
As shown in Fig.~\ref{structure}a, CoTiO$_3$ crystallizes in an ilmenite structure ($R\bar{3}$) that consists of alternating layers of edge-sharing CoO$_6$ or TiO$_6$ octahedra \cite{Newnham1964}. Details of the synthesis and other experimental details are provided in the Supplemental Material.
The magnetic properties of CoTiO$_3$ are determined by the Co$^{2+}$ ions which reside in a slightly buckled honeycomb layer; see Fig.~\ref{structure}b.
The layers are ABC stacked along the $\mathbf{c}$ direction, with neighbouring honeycomb planes displaced diagonally by a third of the
unit cell.

\begin{figure}[tb]
	\centering
\includegraphics[width=0.5\textwidth]{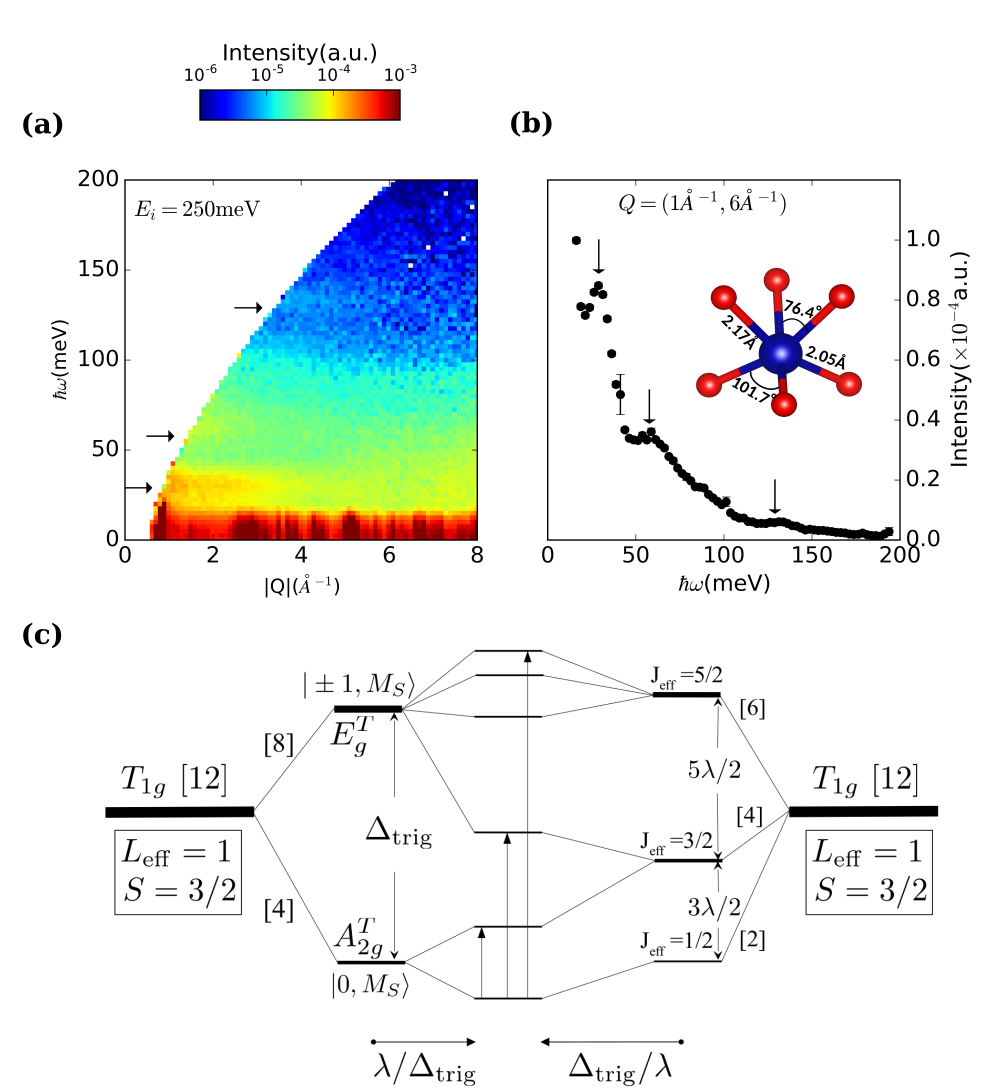}
	\caption{{\bf (a)} Neutron scattering intensity plot on a log scale as a function of energy (in meV) and momentum transfer (in $\AA^{-1}$) for an incident neutron energy of $E_i$=250meV. Since there is no directional dependence of the crystal field excitations, the scattering intensity is averaged over all orientation and only the magnitude of momentum transfer is shown. {\bf (b)} Momentum integrated intensity for $1.0\AA^{-1}<|\mathbf{Q}|<6.0\AA^{-1}$ plotted as a function of energy transfer. Inset showstrigonally distorted oxygen octahedra around each Co$^{2+}$ ion. Horizontal and vertical arrows in (a) and (b), respectively, mark the positions of observed crystal field excitations.
	{\bf (c)} Schematic crystal field levels for Co$^{2+}$ (three holes, total $L_{\rm eff}=1$ and $S=3/2$)
in the regimes $\lambda \ll \Delta_{\rm trig}$ (left)
and $\Delta_{\rm trig} \ll \lambda$ (right). Arrows correspond to sharp transitions seen in (a) and (b).}
	\label{CF}	
\end{figure}

Atomically, each Co$^{2+}$ in a high spin state is surrounded by a trigonally distorted oxygen environment depicted in the inset of Fig.~\ref{CF}b. Local electronic states in Co$^{2+}$ are determined by a combination of trigonal distortion $\Delta_{\rm trig}$ and spin-orbit coupling $\lambda$. To elucidate the magnetic ground state of each Co$^{2+}$ ion, we measured the crystal field excitations using high energy inelastic neutron scattering (INS).
The results shown in Fig.~\ref{CF}a and Fig.~\ref{CF}b clearly reveal three transitions at $\sim30$meV, $\sim$60meV and $\sim$130meV, together with a broad continuum between 60meV-120meV. Their intensities decrease with increasing momentum transfers as shown in Fig.~2a,
which confirms their magnetic origin.

We model the crystal field levels using a single-ion Hamiltonian
$H_{\rm ion} = \lambda \vec L \cdot \vec S + \Delta_{\rm trig} L_z^2$,
where Hund's coupling favors total spin $S\!=\!3/2$ and total orbital angular momentum $L_{\rm eff}\!=\!1$,
and $\hat{z} \! \parallel \! {\bf c}$. We find that $\lambda \! \approx\! 28$~meV
and $\Delta_{\rm trig} \!\approx \! 44$~meV provides a good description of the observed crystal field transitions (see Supplemental Material for details),
leading to transitions at $(29, 58, 108, 117, 129)$\,meV.
The full electronic level diagram of a single Co$^{2+}$ ion determined using these parameters is shown in Fig.~\ref{CF}c.
The three sharp transitions observed in Fig.~\ref{CF}a and Fig.~\ref{CF}b are indicated by vertical arrows, the other two transitions could not be identified unambiguously, but seem to be buried in the broad continuum intensity.

When $\Delta_{\rm trig} \gg \lambda$,
the orbital angular momentum is quenched, $L_z=0$, leading to a pure spin $S=3/2$ moment.
SOC splits
this spin-quartet via an effective single-ion term $H^{\rm eff}_{\rm ion} = \Delta S^2_z$ with $\Delta > 0$,
leading to a ground doublet $S_z \! = \! \pm 1/2$.
In the opposite regime, $\Delta_{\rm trig} \ll \lambda$, SOC will lead a ground $j_{\rm eff} =1/2$ doublet and excited
$j_{\rm eff}=3/2, 5/2$ levels \cite{Liu2018,Sano2018}.
Weak $\Delta_{\rm trig}$ splits the excited levels, and mixes different $j_{\rm eff}$ wavefunctions \cite{Ross2017}.
With increasing $\Delta_{\rm trig}$, the ground $m_j \!=\! \pm1/2$ doublet smoothly connects with the
$S_z \!=\! \pm 1/2$ doublet in the first scenario.
For the fitted values of ($\lambda$, $\Delta_{\rm trig}$), the ratio between $g$-factor parallel ($g_\parallel$) and perpendicular ($g_\perp$) to the honeycomb plane
is $g_\parallel/g_\perp \! \approx\!1.6$, leading to a large anisotropy $(g_\parallel/g_\perp)^2$ in the magnetic susceptibility,
qualitatively consistent with experiment \cite{Wanatabe1980}.
Upon projection to the ground state doublet, even a simple Heisenberg model for the $S\!=\!3/2$ spins leads to a strong easy-plane
exchange anisotropy between $\tilde{S}\!=\!1/2$ pseudospins.

\begin{figure*}[tb]
	\centering
\includegraphics[width=0.8\textwidth]{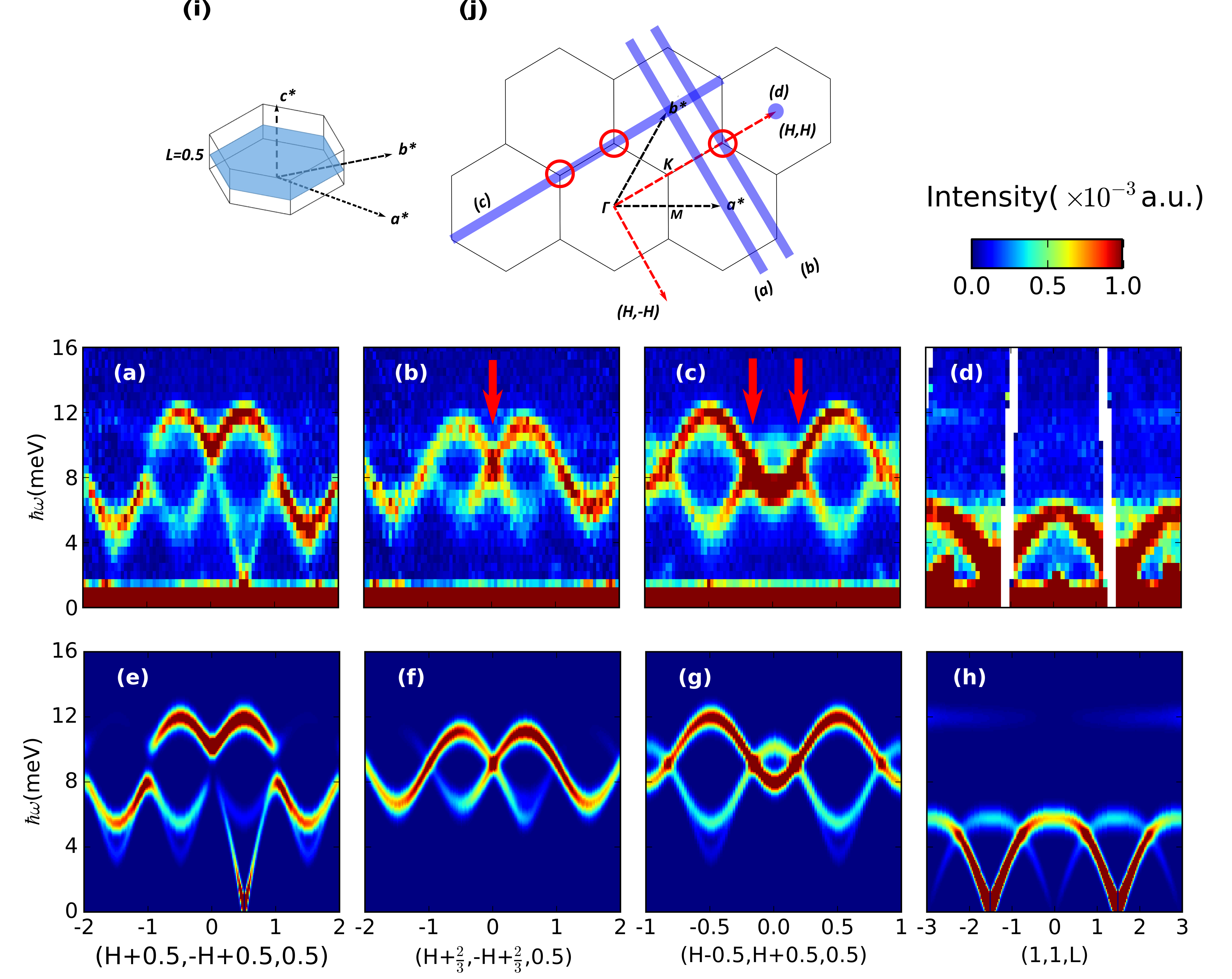}
	\caption{(a)-(d) Momentum and energy resolved neutron scattering intensity map of magnons in CoTiO$_3$. The data was obtained with an incident neutron energy E$_i$=50meV at SEQUOIA at the Spallation Neutron Source (SNS). An arbitrary intensity scale where red (blue) denotes large (small) scattering intensity is used to plot the data. (a)-(c) show magnon excitations along (H,-H) and (H,H) within the honeycomb plane at fixed L=0.5, while (d) shows excitations along L for fixed in plane momentum (1,1). (e)-(h) Calculated magnon spectra using $J_{\parallel, 1}=-4.41$~meV, $J_{\perp,1}=0$meV and $J_{\parallel, 2^\prime=2^{\prime\prime}}=J_{\perp, 2^\prime=2^{\prime\prime}}=0.57$~meV after convolving with the energy experimental resolution of 1meV. (i) Schematics of the 3D BZ and (j) 2D-projection of 3D reciprocal space onto the honeycomb plane. The L=0.5 plane has been shaded in blue in (i). Directions of momentum transfers within the L=0.5 plane in (a) to (c) are denoted with thick blue lines in (j). Intersection of the out-of-plane direction (1,1,L) in (d) with the 2D reciprocal space is shown as a filled blue circle. Red circles in (j) indicate positions of the Dirac point where magnon bands cross, which are denoted by red arrows in (b) and (c).}
	\label{spinwave}	
\end{figure*}

\noindent{\bf Ordered state and magnon dispersion:}
Below the Neel temperature $T_N\approx 38K$, Co$^{2+}$ magnetic moments confined within the $\mathbf{ab}$ plane are ordered ferromagnetically within each honeycomb layer, and antiferromagnetically along the $\mathbf{c}$ direction, giving rise to the ordered structure shown in Fig.~\ref{structure}b\cite{Newnham1964}.
Energy and momentum resolved magnon spectra of CoTiO$_3$ were obtained by inelastic neutron scattering with an incident energy, E$_i$=50meV. In Fig.~\ref{spinwave}, we show magnon spectra for momentum transfers along (H,-H) (Fig.~\ref{spinwave}a,b) and (H,H) (Fig.~\ref{spinwave}c) which lie within the honeycomb plane at fixed L=0.5 as well as along L at (1, 1, L) (Fig.~\ref{spinwave}d). The directions of these momentum transfers have been denoted by thick blue lines in the 2D reciprocal space map in Fig.~\ref{spinwave}j. Strongly dispersive magnon modes extending up to $\sim$12~meV are observed in all directions, which indicates the presence of significant intra-plane and inter-plane couplings in CoTiO$_3$. Magnetic Bragg peaks are located at (1,0,0.5) in Fig.~\ref{spinwave}a and (1,1,$\pm1.5$) in Fig,~\ref{spinwave}d consistent with earlier neutron diffraction results \cite{Newnham1964}. Acoustic magnon modes are found to emanate from these Bragg peaks.
As shown in Fig.~1c, the magnetic unit cell is $\frac{2}{3}$ of the conventional structural unit cell along $\mathbf{c}$; the magnon dispersion should therefore exhibit a periodicity of 1.5 in L. This is inconsistent with the L=3 periodicity apparent in Fig.~2d. As we show below, this reflects the strong XY-nature of magnetic interactions in
CoTiO$_3$.

\begin{figure*}[tb]
	\centering
\includegraphics[width=0.8\textwidth]{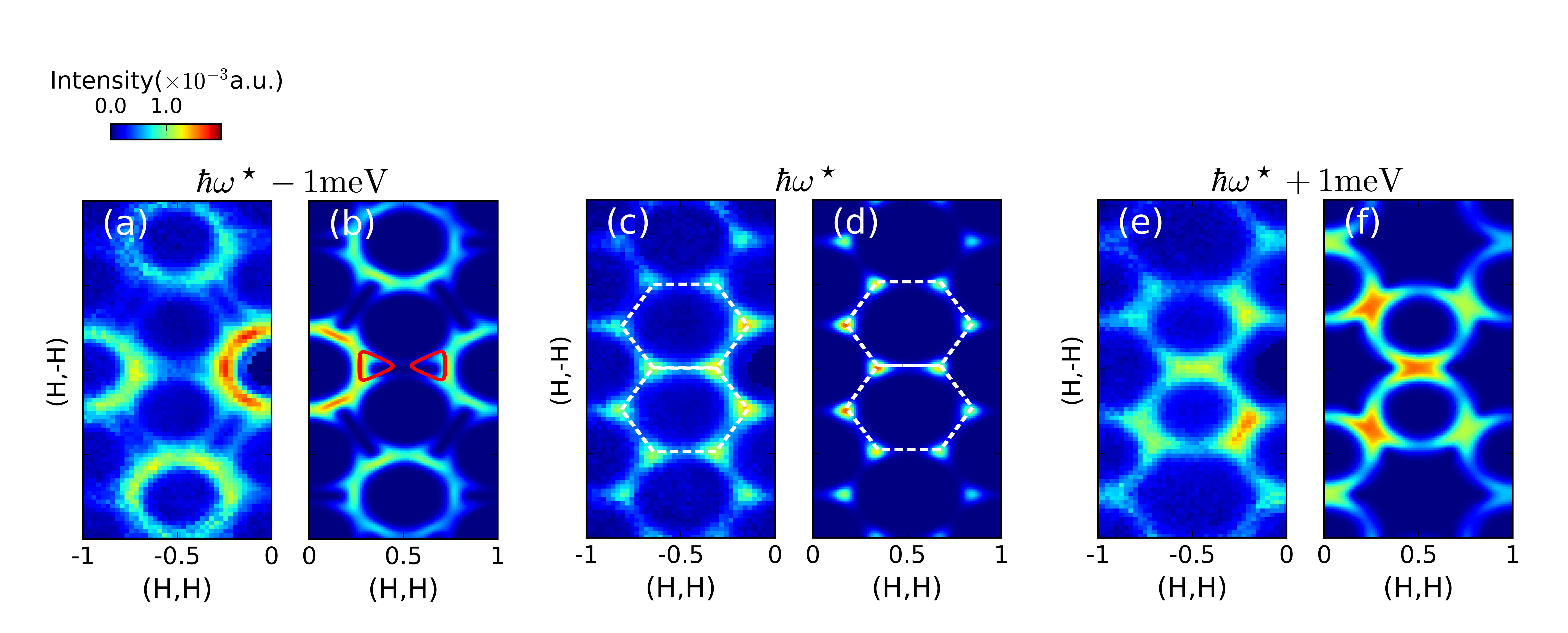}
	\caption{(a,c,e) Constant energy slices of the CoTiO$_3$ magnon spectra in the (H,K) plane. Energies of the slices are chosen to be (a) 1meV below (c) at and (e) 1meV above the Dirac point crossing with $\hbar\omega^\star\sim$8.5meV. Since magnon spectra in CoTiO$_3$ have almost no L dependence at energies close to $\hbar\omega^\star$, all spectra have been integrated over all L measured to improve statistics. (b,d,f) are calculated intensity maps using the same parameters as in Fig.~2. The calculated spectra are convolved with an energy resolution of 1~meV.  An energy integration range of 0.5~meV has been used for both the experimental data and calculation. White dashed lines in (c) and (d) denote the boundaries of 2D BZs. Red triangles in (b) are loops in a constant energy slice of the magnon dispersion below the Dirac points.}
	\label{Eslice}	
\end{figure*}

\begin{figure}[tb]
	\centering
\includegraphics[width=0.5\textwidth]{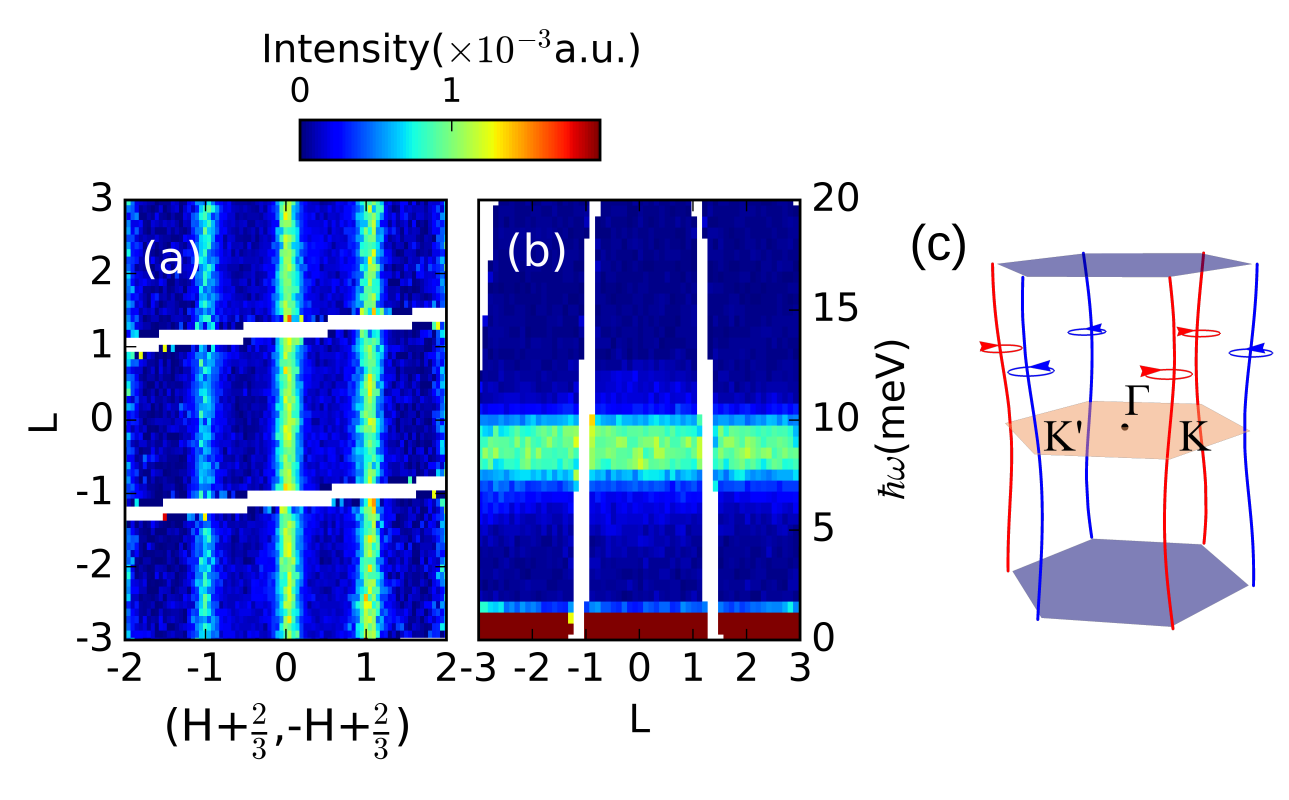}
	\caption{(a) Slice of CoTiO$_3$ magnon spectra in the $(H+\frac{2}{3},-H+\frac{2}{3},L)$ plane at the energy of Dirac point $\hbar\omega^\star$. (b) Magnon dispersion along L with fixed H=K=$\frac{2}{3}$. (c) Schematic plot of Dirac nodal lines in the 3D BZ winding near the corners with additional nearest-neighbor interplane AFM coupling
	$J_{\parallel,3} > 0$ and $J_{\perp,3} < J_{\parallel,3}$.
We depict the vorticity of the complex function ${\cal G}_\bk=A_\bk B_\bk - C_\bk F_\bk$ characterizing the nodal lines  (positive - blue; negative - red). See text for details.}
	\label{DNL}	
\end{figure}

Our data clearly show linear crossings of magnon bands at Dirac points whose positions in reciprocal space are denoted by $\mathbf{K}$ and marked with red circles in Fig.~\ref{spinwave}j. These crossings occur at $\hbar\omega^\star\sim$8.5meV as highlighted by the red arrows in Fig.~\ref{spinwave}b and Fig.~\ref{spinwave}c. The linear dispersions of magnon modes away from these points along (H,-H) and (H,H) are well resolved in our data. These observations unambiguously demonstrate the existence of Dirac magnons in CoTiO$_3$. To visualize Dirac magnons in momentum space, we show slices of magnon dispersions in the (H,K) plane at the energy transfers of $\hbar\omega^\star$ and $\hbar\omega^\star \pm 1$~meV in Fig.~\ref{Eslice}. The constant energy slice at $\hbar\omega^\star$ in Fig.~\ref{Eslice}c shows discrete points at $\mathbf{K}$ that come from crossings of magnon modes. Due to conical dispersions near the Dirac points, a slice of magnon dispersions at an energy below $\hbar\omega^\star$ should show closed loops in momentum space as illustrated in Fig.~\ref{spinwave}b. This is not clear from our data due to finite energy resolution and the fact that the loop traverses three different BZs, each with a very different dynamical structure factor. However, this is clearly captured in our calculation as described below. Remarkably, the Dirac points at different L appear to merge into a nodal line in in the
3D BZ (Fig.~\ref{DNL}a), with no discernible dispersion along $L$ in our data (Fig.~\ref{DNL}b).

{\bf Model Hamiltonian:}
To understand the magnetic excitations in CoTiO$_3$, we carried out linear spin wave calculations using a minimal model containing only nearest neighbour (NN) $\it{intra}$-plane and next nearest neighbour (NNN) $\it{inter}$-plane coupling labelled as 1, $2^{\prime}$ and $2^{\prime\prime}$ in Fig.~\ref{structure}b. The magnitudes of interactions along $2^\prime$ and $2^{\prime\prime}$ are set to be the same due to similar exchange path geometry.
Using $\Delta_{\rm trig}$ and $\lambda$ determined for CoTiO$_3$, the magnetic interactions between the $\tilde{S}$ pseudospins are expected to have strong easy-plane anisotropy
(see Supplemental Material). Interactions between the pseudospins are therefore taken to be of the XXZ-type, given by
$J_\parallel (\tilde{S}_{x,i}\tilde{S}_{x,j}+\tilde{S}_{y,i}\tilde{S}_{y,j})+ J_\perp \tilde{S}_{z,i}\tilde{S}_{z,j}$.

The neutron data shown in Figs. ~\ref{spinwave} and Fig.~\ref{Eslice} are described very well by this simple model Hamiltonian with just two parameters: a ferromagnetic XY exchange coupling along bond 1 shown in Fig. 1c   ($J_{\parallel,1}=-4.4(9)$~meV and $J_{\perp,1}$=0~meV), and a Heisenberg antiferromagnetic exhange interaction on bonds $2^{\prime}$ and $2^{\prime\prime}$ ($J_{\parallel,2^\prime=2^{\prime\prime}}
=J_{\perp,2^\prime=2^{\prime\prime}}=0.6(1)$~meV).
We note that the fitting results are not changed by adding a small NN inter-plane coupling ($\lesssim$1~meV),
which connects two Co$^{2+}$ directly on top of each other along the $\mathbf{c}$ axis. The effect of such a NN inter-plane
coupling on the magnon dispersion is likely to be negligible compared to $2^\prime$ and $2^{\prime\prime}$ interactions because each spin
has only one out-of-plane NN but 9 NNN's. To compare our calculation directly with the data, we calculate the magnetic inelastic intensity using the expression $I(\mathbf{Q},\omega)\propto \frac{1}{2}g_\parallel^2\mathcal{S}_\parallel+g_\perp^2\mathcal{S}_\perp$ where $\mathcal{S}_\parallel$ ($\mathcal{S}_\perp$) denotes the in-plane (out-of-plane) fluctuations of pseudospins. (See Supplemental Material for derivation of this expression and further fitting details.) Magnon spectra calculated within the linear spin wave theory is shown in Fig.~\ref{spinwave}e-h. Our calculation gives four magnon modes due to the presence of four spins in a primitive unit cell. The scattering intensity in CoTiO$_3$ mostly comes from the two modes contributing to $\mathcal{S}_\parallel$ because of the larger $g_\parallel$ as well as stronger in-plane spin fluctuations. This provides a natural explanation for the apparent $L=3$ periodicity in Fig.~\ref{spinwave}d. The prominent magnon mode emanating from $L=\pm 1.5$ is due to the $\mathcal{S}_\parallel$ contribution, while the much weaker $\mathcal{S}_\perp$ contribution is shifted by 1.5 along $L$ with respect to $\mathcal{S}_\parallel$. The overall periodicity in $L$ is therefore 1.5, but the dominant intensity of $\mathcal{S}_\parallel$ makes it difficult to see the true periodicity.
The weak magnon intensity due to $\mathcal{S}_\perp$ can be identified in the data in Fig.~\ref{spinwave}a and Fig.~\ref{spinwave}d, in support of our effective spin model.

The calculated magnon spectra in Fig.~\ref{spinwave}f and Fig.~\ref{spinwave}g clearly show crossings of magnon modes at the $\mathbf{K}$ points, which is consistent with results shown in Fig.~\ref{spinwave}b and c. By taking into account finite energy resolution, constant energy slices in the vicinity of $\hbar\omega^\star$ determined from our model (Fig.~\ref{Eslice}b,d,f) are in good agreement with Fig.~\ref{Eslice}a,c,e. In particular, the triangular loops in the constant energy slice below $\hbar\omega^\star$ broaden into two lines connecting bigger rings (Fig.~\ref{Eslice}b), entirely consistent with our data in Fig.~\ref{Eslice}a.

Moreover, our model reproduces the nodal line of magnon modes crossings along L which intersects the 2D BZ at $\mathbf{K}$. The nodal line observed in CoTiO$_3$ is a magnon analogue of the Dirac Nodal Line (DNL) in 3D semimetals \cite{topologicalphase_review}, or the symmetry protected line degeneracy of 3D electronic bands.
Similar to
Dirac nodal lines in 3D electronic materials without SOC, the
XXZ spin model for CoTiO$_3$ features a topological Berry phase winding which ensures the stability of the DNL observed in our data even when such
perturbations due to weak further neighbour interactions are included (see Supplemental Material for further discussion).
In the presence of significant nearest-neighbour interplane XXZ exchange (denoted by $J_3$),
the DNL is no longer pinned at $\mathbf{K}$ for an arbitrary L ,
but starts to spiral around it. Within our momentum resolution,
we could not resolve any deviation of the DNL away from the $\mathbf{K}$ points, which justifies the assumption of ignoring further-neighbour
interactions in our model and linear spin wave calculations. Furthermore, given the in-plane magnetic order, the impact of the SOC induced
Dzyaloshinskii-Moriya interaction also vanishes at this high symmetry point by virtue of $C_3$ symmetry.
Since relative magnitudes between magnetic interactions can be tuned by changing the lattice structure, one might be able to drive a systematic movement of the
DNL in CoTiO$_3$ by an applied pressure.
Such DNLs may be also found in other ilmenite magnets, e.g., FeTiO$_3$ \cite{Kato1986}; however, the location of DNL might vary depending on their
anisotropies \cite{Goodenough1967}.

{\bf Beyond the XXZ model:}
One important observation which is not captured by our XXZ model is the existence of a small magnetic anisotropy within the honeycomb plane. This is inferred from
the highly non-linear magnetization at $T\!=\! 5$\,K for in-plane fields $\sim\! 1$-$4$ Tesla, with a peak susceptibility at $\sim \! 2$\,Tesla, which is likely a result of rotation of magnetic domains \cite{balbashov2017}. An in-plane anisotropy also implies the existence of a small gap at the magnetic zone center in the magnon dispersion. Although our experimental resolution does not allow us to determine the gap
size directly, extrapolation of the magnon dispersion in Fig.~3a suggests a gap of order $\sim\! 1$\,meV (See Supplemental Material). Such a gap to the
Goldstone mode can arise from bond-anisotropic exchange couplings, like the Kitaev interaction, due to quantum order by disorder which pins the order parameter to the
crystal axes. A phenomenological way to account for this is via a pinning field $ (-1)^z g_\parallel \mu_B h \tilde{S}_x$, staggered from layer to layer, deep in the ordered
phase. We find that incorporating a pinning field $\sim\!2$\,Tesla, based on the magnetization data,
leads to a $\sim\! 1$\,meV zone center gap in the magnon dispersion consistent with the above INS estimate.
The Hamiltonian including such a pinning field continues to support a DNL near the $\mathbf{K}$ points. A complete theoretical study of such weak bond-anisotropic
exchanges will be discussed in a separate publication \cite{Khait2019}.

{\bf Conclusion:}
In conclusion, we show that the ilmenite antiferromagnet CoTiO$_3$ is a magnon analogue of ABC stacked graphene. Our identification of the Dirac magnon phase, and possibly a magnon analogue of a Dirac Nodal Line, in CoTiO$_3$ serves as a starting point to study transitions into other topological phases such as 3D Weyl magnons and magnon topological insulators. The ease of chemical substitution on both Co and Ti sites and high mechanical strength of CoTiO$_3$ make it an ideal material for future studies of the impact of
doping, hydrostatic pressure, strain, and magnetic fields, on gapless Dirac magnons.
The close resemblance between CoTiO$_3$ and graphene allows a direct comparison between bosonic and fermionic responses to external perturbations.
For example, the Dirac magnons in CoTiO$_3$ are an ideal model system to study emergent gauge fields through strain engineering \cite{Guinea2009}, or renormalization of the magnon bands resulting from the inter-particle interaction between Dirac bosons \cite{Yelon1971,Pershoguba2018}.
Finally,  unlike a simple Heisenberg ferromagnet, for which the electronic analogue is a simple
tight-binding hopping Hamiltonian, the full model Hamiltonian for CoTiO$_3$ contains
pairing terms analogous to the Bogoliubov de-Gennes Hamiltonian  \cite{topologicalphase_review} of a superconductor.
Such unconventional superconducting phases on the layered honeycomb lattice have begun to garner great
attention due to the recent discovery of superconductivity in `magic angle’' twisted bilayer graphene \cite{Cao2018}, providing further impetus to explore
such remarkable analogies between electronic quasiparticles and bosonic magnons.

\begin{acknowledgments}
Work at the University of Toronto was supported by the
Natural Science and Engineering Research Council (NSERC)
of Canada. GJS acknowledges the support provided by MOST-Taiwan under project number 105-2112-M-027-003-MY3.
This research used resources at the Spallation Neutron Source, a DOE Office of Science User Facility operated by the Oak Ridge National Laboratory. Use of the MAD beamline at the McMaster Nuclear Reactor is supported by McMaster University and the Canada Foundation for Innovation.

\end{acknowledgments}


\begin{thebibliography}{33}%
\makeatletter
\providecommand \@ifxundefined [1]{%
 \@ifx{#1\undefined}
}%
\providecommand \@ifnum [1]{%
 \ifnum #1\expandafter \@firstoftwo
 \else \expandafter \@secondoftwo
 \fi
}%
\providecommand \@ifx [1]{%
 \ifx #1\expandafter \@firstoftwo
 \else \expandafter \@secondoftwo
 \fi
}%
\providecommand \natexlab [1]{#1}%
\providecommand \enquote  [1]{``#1''}%
\providecommand \bibnamefont  [1]{#1}%
\providecommand \bibfnamefont [1]{#1}%
\providecommand \citenamefont [1]{#1}%
\providecommand \href@noop [0]{\@secondoftwo}%
\providecommand \href [0]{\begingroup \@sanitize@url \@href}%
\providecommand \@href[1]{\@@startlink{#1}\@@href}%
\providecommand \@@href[1]{\endgroup#1\@@endlink}%
\providecommand \@sanitize@url [0]{\catcode `\\12\catcode `\$12\catcode
  `\&12\catcode `\#12\catcode `\^12\catcode `\_12\catcode `\%12\relax}%
\providecommand \@@startlink[1]{}%
\providecommand \@@endlink[0]{}%
\providecommand \url  [0]{\begingroup\@sanitize@url \@url }%
\providecommand \@url [1]{\endgroup\@href {#1}{\urlprefix }}%
\providecommand \urlprefix  [0]{URL }%
\providecommand \Eprint [0]{\href }%
\providecommand \doibase [0]{http://dx.doi.org/}%
\providecommand \selectlanguage [0]{\@gobble}%
\providecommand \bibinfo  [0]{\@secondoftwo}%
\providecommand \bibfield  [0]{\@secondoftwo}%
\providecommand \translation [1]{[#1]}%
\providecommand \BibitemOpen [0]{}%
\providecommand \bibitemStop [0]{}%
\providecommand \bibitemNoStop [0]{.\EOS\space}%
\providecommand \EOS [0]{\spacefactor3000\relax}%
\providecommand \BibitemShut  [1]{\csname bibitem#1\endcsname}%
\let\auto@bib@innerbib\@empty
\bibitem [{\citenamefont {{Novoselov}}\ \emph {et~al.}(2005)\citenamefont
  {{Novoselov}}, \citenamefont {{Geim}}, \citenamefont {{Morozov}},
  \citenamefont {{Jiang}}, \citenamefont {{Katsnelson}}, \citenamefont
  {{Grigorieva}}, \citenamefont {{Dubonos}},\ and\ \citenamefont
  {{Firsov}}}]{Novoselov2005}%
  \BibitemOpen
  \bibfield  {author} {\bibinfo {author} {\bibfnamefont {K.~S.}\ \bibnamefont
  {{Novoselov}}}, \bibinfo {author} {\bibfnamefont {A.~K.}\ \bibnamefont
  {{Geim}}}, \bibinfo {author} {\bibfnamefont {S.~V.}\ \bibnamefont
  {{Morozov}}}, \bibinfo {author} {\bibfnamefont {D.}~\bibnamefont {{Jiang}}},
  \bibinfo {author} {\bibfnamefont {M.~I.}\ \bibnamefont {{Katsnelson}}},
  \bibinfo {author} {\bibfnamefont {I.~V.}\ \bibnamefont {{Grigorieva}}},
  \bibinfo {author} {\bibfnamefont {S.~V.}\ \bibnamefont {{Dubonos}}}, \ and\
  \bibinfo {author} {\bibfnamefont {A.~A.}\ \bibnamefont {{Firsov}}},\
  }\bibfield  {title} {\enquote {\bibinfo {title} {{Two-dimensional gas of
  massless Dirac fermions in graphene}},}\ }\href {\doibase
  10.1038/nature04233} {\bibfield  {journal} {\bibinfo  {journal} {Nature}\
  }\textbf {\bibinfo {volume} {438}},\ \bibinfo {pages} {197--200} (\bibinfo
  {year} {2005})},\ \Eprint {http://arxiv.org/abs/cond-mat/0509330}
  {cond-mat/0509330} \BibitemShut {NoStop}%
\bibitem [{\citenamefont {{Zhang}}\ \emph {et~al.}(2005)\citenamefont
  {{Zhang}}, \citenamefont {{Tan}}, \citenamefont {{Stormer}},\ and\
  \citenamefont {{Kim}}}]{Zhang2005}%
  \BibitemOpen
  \bibfield  {author} {\bibinfo {author} {\bibfnamefont {Y.}~\bibnamefont
  {{Zhang}}}, \bibinfo {author} {\bibfnamefont {Y.-W.}\ \bibnamefont {{Tan}}},
  \bibinfo {author} {\bibfnamefont {H.~L.}\ \bibnamefont {{Stormer}}}, \ and\
  \bibinfo {author} {\bibfnamefont {P.}~\bibnamefont {{Kim}}},\ }\bibfield
  {title} {\enquote {\bibinfo {title} {{Experimental observation of the quantum
  Hall effect and Berry's phase in graphene}},}\ }\href {\doibase
  10.1038/nature04235} {\bibfield  {journal} {\bibinfo  {journal} {Nature}\
  }\textbf {\bibinfo {volume} {438}},\ \bibinfo {pages} {201--204} (\bibinfo
  {year} {2005})},\ \Eprint {http://arxiv.org/abs/cond-mat/0509355}
  {cond-mat/0509355} \BibitemShut {NoStop}%
\bibitem [{\citenamefont {Katsnelson}\ \emph {et~al.}(2006)\citenamefont
  {Katsnelson}, \citenamefont {Novoselov},\ and\ \citenamefont
  {Geim}}]{Katsnelson2006}%
  \BibitemOpen
  \bibfield  {author} {\bibinfo {author} {\bibfnamefont {M.~I.}\ \bibnamefont
  {Katsnelson}}, \bibinfo {author} {\bibfnamefont {K.~S.}\ \bibnamefont
  {Novoselov}}, \ and\ \bibinfo {author} {\bibfnamefont {A.~K.}\ \bibnamefont
  {Geim}},\ }\bibfield  {title} {\enquote {\bibinfo {title} {Chiral tunnelling
  and the klein paradox in graphene},}\ }\href {\doibase 10.1038/nphys384}
  {\bibfield  {journal} {\bibinfo  {journal} {Nature Physics}\ }\textbf
  {\bibinfo {volume} {2}},\ \bibinfo {pages} {620} (\bibinfo {year}
  {2006})}\BibitemShut {NoStop}%
\bibitem [{\citenamefont {Castro~Neto}\ \emph {et~al.}(2009)\citenamefont
  {Castro~Neto}, \citenamefont {Guinea}, \citenamefont {Peres}, \citenamefont
  {Novoselov},\ and\ \citenamefont {Geim}}]{graphene_review}%
  \BibitemOpen
  \bibfield  {author} {\bibinfo {author} {\bibfnamefont {A.~H.}\ \bibnamefont
  {Castro~Neto}}, \bibinfo {author} {\bibfnamefont {F.}~\bibnamefont {Guinea}},
  \bibinfo {author} {\bibfnamefont {N.~M.~R.}\ \bibnamefont {Peres}}, \bibinfo
  {author} {\bibfnamefont {K.~S.}\ \bibnamefont {Novoselov}}, \ and\ \bibinfo
  {author} {\bibfnamefont {A.~K.}\ \bibnamefont {Geim}},\ }\bibfield  {title}
  {\enquote {\bibinfo {title} {The electronic properties of graphene},}\ }\href
  {\doibase 10.1103/RevModPhys.81.109} {\bibfield  {journal} {\bibinfo
  {journal} {Rev. Mod. Phys.}\ }\textbf {\bibinfo {volume} {81}},\ \bibinfo
  {pages} {109--162} (\bibinfo {year} {2009})}\BibitemShut {NoStop}%
\bibitem [{\citenamefont {Wehling}\ \emph {et~al.}(2014)\citenamefont
  {Wehling}, \citenamefont {Black-Schaffer},\ and\ \citenamefont
  {Balatsky}}]{Wehling2014}%
  \BibitemOpen
  \bibfield  {author} {\bibinfo {author} {\bibfnamefont {T.O.}\ \bibnamefont
  {Wehling}}, \bibinfo {author} {\bibfnamefont {A.M.}\ \bibnamefont
  {Black-Schaffer}}, \ and\ \bibinfo {author} {\bibfnamefont {A.V.}\
  \bibnamefont {Balatsky}},\ }\bibfield  {title} {\enquote {\bibinfo {title}
  {Dirac materials},}\ }\href {\doibase 10.1080/00018732.2014.927109}
  {\bibfield  {journal} {\bibinfo  {journal} {Advances in Physics}\ }\textbf
  {\bibinfo {volume} {63}},\ \bibinfo {pages} {1--76} (\bibinfo {year}
  {2014})},\ \Eprint
  {http://arxiv.org/abs/https://doi.org/10.1080/00018732.2014.927109}
  {https://doi.org/10.1080/00018732.2014.927109} \BibitemShut {NoStop}%
\bibitem [{\citenamefont {Chiu}\ \emph {et~al.}(2016)\citenamefont {Chiu},
  \citenamefont {Teo}, \citenamefont {Schnyder},\ and\ \citenamefont
  {Ryu}}]{topologicalphase_review}%
  \BibitemOpen
  \bibfield  {author} {\bibinfo {author} {\bibfnamefont {Ching-Kai}\
  \bibnamefont {Chiu}}, \bibinfo {author} {\bibfnamefont {Jeffrey C.~Y.}\
  \bibnamefont {Teo}}, \bibinfo {author} {\bibfnamefont {Andreas~P.}\
  \bibnamefont {Schnyder}}, \ and\ \bibinfo {author} {\bibfnamefont {Shinsei}\
  \bibnamefont {Ryu}},\ }\bibfield  {title} {\enquote {\bibinfo {title}
  {Classification of topological quantum matter with symmetries},}\ }\href
  {\doibase 10.1103/RevModPhys.88.035005} {\bibfield  {journal} {\bibinfo
  {journal} {Rev. Mod. Phys.}\ }\textbf {\bibinfo {volume} {88}},\ \bibinfo
  {pages} {035005} (\bibinfo {year} {2016})}\BibitemShut {NoStop}%
\bibitem [{\citenamefont {Lu}\ \emph {et~al.}(2014)\citenamefont {Lu},
  \citenamefont {Joannopoulos},\ and\ \citenamefont {Soljacic}}]{Lu2014}%
  \BibitemOpen
  \bibfield  {author} {\bibinfo {author} {\bibfnamefont {Ling}\ \bibnamefont
  {Lu}}, \bibinfo {author} {\bibfnamefont {John~D.}\ \bibnamefont
  {Joannopoulos}}, \ and\ \bibinfo {author} {\bibfnamefont {Marin}\
  \bibnamefont {Soljacic}},\ }\bibfield  {title} {\enquote {\bibinfo {title}
  {Topological photonics},}\ }\href {\doibase 10.1038/nphoton.2014.248}
  {\bibfield  {journal} {\bibinfo  {journal} {Nature Photonics}\ }\textbf
  {\bibinfo {volume} {8}},\ \bibinfo {pages} {821} (\bibinfo {year}
  {2014})}\BibitemShut {NoStop}%
\bibitem [{\citenamefont {Ozawa}\ \emph {et~al.}(2019)\citenamefont {Ozawa},
  \citenamefont {Price}, \citenamefont {Amo}, \citenamefont {Goldman},
  \citenamefont {Hafezi}, \citenamefont {Lu}, \citenamefont {Rechtsman},
  \citenamefont {Schuster}, \citenamefont {Simon}, \citenamefont {Zilberberg},\
  and\ \citenamefont {Carusotto}}]{topologicalphoton_review}%
  \BibitemOpen
  \bibfield  {author} {\bibinfo {author} {\bibfnamefont {Tomoki}\ \bibnamefont
  {Ozawa}}, \bibinfo {author} {\bibfnamefont {Hannah~M.}\ \bibnamefont
  {Price}}, \bibinfo {author} {\bibfnamefont {Alberto}\ \bibnamefont {Amo}},
  \bibinfo {author} {\bibfnamefont {Nathan}\ \bibnamefont {Goldman}}, \bibinfo
  {author} {\bibfnamefont {Mohammad}\ \bibnamefont {Hafezi}}, \bibinfo {author}
  {\bibfnamefont {Ling}\ \bibnamefont {Lu}}, \bibinfo {author} {\bibfnamefont
  {Mikael~C.}\ \bibnamefont {Rechtsman}}, \bibinfo {author} {\bibfnamefont
  {David}\ \bibnamefont {Schuster}}, \bibinfo {author} {\bibfnamefont
  {Jonathan}\ \bibnamefont {Simon}}, \bibinfo {author} {\bibfnamefont {Oded}\
  \bibnamefont {Zilberberg}}, \ and\ \bibinfo {author} {\bibfnamefont {Iacopo}\
  \bibnamefont {Carusotto}},\ }\bibfield  {title} {\enquote {\bibinfo {title}
  {Topological photonics},}\ }\href {\doibase 10.1103/RevModPhys.91.015006}
  {\bibfield  {journal} {\bibinfo  {journal} {Rev. Mod. Phys.}\ }\textbf
  {\bibinfo {volume} {91}},\ \bibinfo {pages} {015006} (\bibinfo {year}
  {2019})}\BibitemShut {NoStop}%
\bibitem [{\citenamefont {{Jin}}\ \emph {et~al.}(2018)\citenamefont {{Jin}},
  \citenamefont {{Wang}},\ and\ \citenamefont {{Xu}}}]{Jin2018}%
  \BibitemOpen
  \bibfield  {author} {\bibinfo {author} {\bibfnamefont {Yuanjun}\ \bibnamefont
  {{Jin}}}, \bibinfo {author} {\bibfnamefont {Rui}\ \bibnamefont {{Wang}}}, \
  and\ \bibinfo {author} {\bibfnamefont {Hu}~\bibnamefont {{Xu}}},\ }\bibfield
  {title} {\enquote {\bibinfo {title} {{Recipe for Dirac Phonon States with a
  Quantized Valley Berry Phase in Two-Dimensional Hexagonal Lattices}},}\
  }\href {\doibase 10.1021/acs.nanolett.8b03492} {\bibfield  {journal}
  {\bibinfo  {journal} {Nano Letters}\ }\textbf {\bibinfo {volume} {18}},\
  \bibinfo {pages} {7755--7760} (\bibinfo {year} {2018})},\ \Eprint
  {http://arxiv.org/abs/1811.09492} {arXiv:1811.09492 [cond-mat.mes-hall]}
  \BibitemShut {NoStop}%
\bibitem [{\citenamefont {Fransson}\ \emph {et~al.}(2016)\citenamefont
  {Fransson}, \citenamefont {Black-Schaffer},\ and\ \citenamefont
  {Balatsky}}]{Fransson2016}%
  \BibitemOpen
  \bibfield  {author} {\bibinfo {author} {\bibfnamefont {J.}~\bibnamefont
  {Fransson}}, \bibinfo {author} {\bibfnamefont {A.~M.}\ \bibnamefont
  {Black-Schaffer}}, \ and\ \bibinfo {author} {\bibfnamefont {A.~V.}\
  \bibnamefont {Balatsky}},\ }\bibfield  {title} {\enquote {\bibinfo {title}
  {Magnon dirac materials},}\ }\href {\doibase 10.1103/PhysRevB.94.075401}
  {\bibfield  {journal} {\bibinfo  {journal} {Phys. Rev. B}\ }\textbf {\bibinfo
  {volume} {94}},\ \bibinfo {pages} {075401} (\bibinfo {year}
  {2016})}\BibitemShut {NoStop}%
\bibitem [{\citenamefont {Owerre}(2016)}]{Owerre2016}%
  \BibitemOpen
  \bibfield  {author} {\bibinfo {author} {\bibfnamefont {Solomon}\ \bibnamefont
  {Owerre}},\ }\bibfield  {title} {\enquote {\bibinfo {title} {A first
  theoretical realization of honeycomb topological magnon insulator},}\ }\href
  {\doibase 10.1088/0953-8984/28/38/386001} {\bibfield  {journal} {\bibinfo
  {journal} {Journal of Physics: Condensed Matter}\ }\textbf {\bibinfo {volume}
  {28}},\ \bibinfo {pages} {386001} (\bibinfo {year} {2016})}\BibitemShut
  {NoStop}%
\bibitem [{\citenamefont {Pershoguba}\ \emph {et~al.}(2018)\citenamefont
  {Pershoguba}, \citenamefont {Banerjee}, \citenamefont {Lashley},
  \citenamefont {Park}, \citenamefont {\AA{}gren}, \citenamefont {Aeppli},\
  and\ \citenamefont {Balatsky}}]{Pershoguba2018}%
  \BibitemOpen
  \bibfield  {author} {\bibinfo {author} {\bibfnamefont {Sergey~S.}\
  \bibnamefont {Pershoguba}}, \bibinfo {author} {\bibfnamefont {Saikat}\
  \bibnamefont {Banerjee}}, \bibinfo {author} {\bibfnamefont {J.~C.}\
  \bibnamefont {Lashley}}, \bibinfo {author} {\bibfnamefont {Jihwey}\
  \bibnamefont {Park}}, \bibinfo {author} {\bibfnamefont {Hans}\ \bibnamefont
  {\AA{}gren}}, \bibinfo {author} {\bibfnamefont {Gabriel}\ \bibnamefont
  {Aeppli}}, \ and\ \bibinfo {author} {\bibfnamefont {Alexander~V.}\
  \bibnamefont {Balatsky}},\ }\bibfield  {title} {\enquote {\bibinfo {title}
  {Dirac magnons in honeycomb ferromagnets},}\ }\href {\doibase
  10.1103/PhysRevX.8.011010} {\bibfield  {journal} {\bibinfo  {journal} {Phys.
  Rev. X}\ }\textbf {\bibinfo {volume} {8}},\ \bibinfo {pages} {011010}
  (\bibinfo {year} {2018})}\BibitemShut {NoStop}%
\bibitem [{\citenamefont {Boyko}\ \emph {et~al.}(2018)\citenamefont {Boyko},
  \citenamefont {Balatsky},\ and\ \citenamefont {Haraldsen}}]{Boyko2018}%
  \BibitemOpen
  \bibfield  {author} {\bibinfo {author} {\bibfnamefont {D.}~\bibnamefont
  {Boyko}}, \bibinfo {author} {\bibfnamefont {A.~V.}\ \bibnamefont {Balatsky}},
  \ and\ \bibinfo {author} {\bibfnamefont {J.~T.}\ \bibnamefont {Haraldsen}},\
  }\bibfield  {title} {\enquote {\bibinfo {title} {Evolution of magnetic dirac
  bosons in a honeycomb lattice},}\ }\href {\doibase
  10.1103/PhysRevB.97.014433} {\bibfield  {journal} {\bibinfo  {journal} {Phys.
  Rev. B}\ }\textbf {\bibinfo {volume} {97}},\ \bibinfo {pages} {014433}
  (\bibinfo {year} {2018})}\BibitemShut {NoStop}%
\bibitem [{\citenamefont {Zhang}\ \emph {et~al.}(2013)\citenamefont {Zhang},
  \citenamefont {Ren}, \citenamefont {Wang},\ and\ \citenamefont
  {Li}}]{Zhang2013}%
  \BibitemOpen
  \bibfield  {author} {\bibinfo {author} {\bibfnamefont {Lifa}\ \bibnamefont
  {Zhang}}, \bibinfo {author} {\bibfnamefont {Jie}\ \bibnamefont {Ren}},
  \bibinfo {author} {\bibfnamefont {Jian-Sheng}\ \bibnamefont {Wang}}, \ and\
  \bibinfo {author} {\bibfnamefont {Baowen}\ \bibnamefont {Li}},\ }\bibfield
  {title} {\enquote {\bibinfo {title} {Topological magnon insulator in
  insulating ferromagnet},}\ }\href {\doibase 10.1103/PhysRevB.87.144101}
  {\bibfield  {journal} {\bibinfo  {journal} {Phys. Rev. B}\ }\textbf {\bibinfo
  {volume} {87}},\ \bibinfo {pages} {144101} (\bibinfo {year}
  {2013})}\BibitemShut {NoStop}%
\bibitem [{\citenamefont {Mook}\ \emph {et~al.}(2014)\citenamefont {Mook},
  \citenamefont {Henk},\ and\ \citenamefont {Mertig}}]{Mook2014}%
  \BibitemOpen
  \bibfield  {author} {\bibinfo {author} {\bibfnamefont {Alexander}\
  \bibnamefont {Mook}}, \bibinfo {author} {\bibfnamefont {J\"urgen}\
  \bibnamefont {Henk}}, \ and\ \bibinfo {author} {\bibfnamefont {Ingrid}\
  \bibnamefont {Mertig}},\ }\bibfield  {title} {\enquote {\bibinfo {title}
  {Edge states in topological magnon insulators},}\ }\href {\doibase
  10.1103/PhysRevB.90.024412} {\bibfield  {journal} {\bibinfo  {journal} {Phys.
  Rev. B}\ }\textbf {\bibinfo {volume} {90}},\ \bibinfo {pages} {024412}
  (\bibinfo {year} {2014})}\BibitemShut {NoStop}%
\bibitem [{\citenamefont {Chisnell}\ \emph {et~al.}(2015)\citenamefont
  {Chisnell}, \citenamefont {Helton}, \citenamefont {Freedman}, \citenamefont
  {Singh}, \citenamefont {Bewley}, \citenamefont {Nocera},\ and\ \citenamefont
  {Lee}}]{Chisnell2015}%
  \BibitemOpen
  \bibfield  {author} {\bibinfo {author} {\bibfnamefont {R.}~\bibnamefont
  {Chisnell}}, \bibinfo {author} {\bibfnamefont {J.~S.}\ \bibnamefont
  {Helton}}, \bibinfo {author} {\bibfnamefont {D.~E.}\ \bibnamefont
  {Freedman}}, \bibinfo {author} {\bibfnamefont {D.~K.}\ \bibnamefont {Singh}},
  \bibinfo {author} {\bibfnamefont {R.~I.}\ \bibnamefont {Bewley}}, \bibinfo
  {author} {\bibfnamefont {D.~G.}\ \bibnamefont {Nocera}}, \ and\ \bibinfo
  {author} {\bibfnamefont {Y.~S.}\ \bibnamefont {Lee}},\ }\bibfield  {title}
  {\enquote {\bibinfo {title} {Topological magnon bands in a kagome lattice
  ferromagnet},}\ }\href {\doibase 10.1103/PhysRevLett.115.147201} {\bibfield
  {journal} {\bibinfo  {journal} {Phys. Rev. Lett.}\ }\textbf {\bibinfo
  {volume} {115}},\ \bibinfo {pages} {147201} (\bibinfo {year}
  {2015})}\BibitemShut {NoStop}%
\bibitem [{\citenamefont {Chen}\ \emph {et~al.}(2018)\citenamefont {Chen},
  \citenamefont {Chung}, \citenamefont {Gao}, \citenamefont {Chen},
  \citenamefont {Stone}, \citenamefont {Kolesnikov}, \citenamefont {Huang},\
  and\ \citenamefont {Dai}}]{Chen2018}%
  \BibitemOpen
  \bibfield  {author} {\bibinfo {author} {\bibfnamefont {Lebing}\ \bibnamefont
  {Chen}}, \bibinfo {author} {\bibfnamefont {Jae-Ho}\ \bibnamefont {Chung}},
  \bibinfo {author} {\bibfnamefont {Bin}\ \bibnamefont {Gao}}, \bibinfo
  {author} {\bibfnamefont {Tong}\ \bibnamefont {Chen}}, \bibinfo {author}
  {\bibfnamefont {Matthew~B.}\ \bibnamefont {Stone}}, \bibinfo {author}
  {\bibfnamefont {Alexander~I.}\ \bibnamefont {Kolesnikov}}, \bibinfo {author}
  {\bibfnamefont {Qingzhen}\ \bibnamefont {Huang}}, \ and\ \bibinfo {author}
  {\bibfnamefont {Pengcheng}\ \bibnamefont {Dai}},\ }\bibfield  {title}
  {\enquote {\bibinfo {title} {Topological spin excitations in honeycomb
  ferromagnet ${\mathrm{cri}}_{3}$},}\ }\href {\doibase
  10.1103/PhysRevX.8.041028} {\bibfield  {journal} {\bibinfo  {journal} {Phys.
  Rev. X}\ }\textbf {\bibinfo {volume} {8}},\ \bibinfo {pages} {041028}
  (\bibinfo {year} {2018})}\BibitemShut {NoStop}%
\bibitem [{\citenamefont {{Yao}}\ \emph {et~al.}(2018)\citenamefont {{Yao}},
  \citenamefont {{Li}}, \citenamefont {{Wang}}, \citenamefont {{Xue}},
  \citenamefont {{Dan}}, \citenamefont {{Iida}}, \citenamefont {{Kamazawa}},
  \citenamefont {{Li}}, \citenamefont {{Fang}},\ and\ \citenamefont
  {{Li}}}]{Yao2018}%
  \BibitemOpen
  \bibfield  {author} {\bibinfo {author} {\bibfnamefont {Weiliang}\
  \bibnamefont {{Yao}}}, \bibinfo {author} {\bibfnamefont {Chenyuan}\
  \bibnamefont {{Li}}}, \bibinfo {author} {\bibfnamefont {Lichen}\ \bibnamefont
  {{Wang}}}, \bibinfo {author} {\bibfnamefont {Shangjie}\ \bibnamefont
  {{Xue}}}, \bibinfo {author} {\bibfnamefont {Yang}\ \bibnamefont {{Dan}}},
  \bibinfo {author} {\bibfnamefont {Kazuki}\ \bibnamefont {{Iida}}}, \bibinfo
  {author} {\bibfnamefont {Kazuya}\ \bibnamefont {{Kamazawa}}}, \bibinfo
  {author} {\bibfnamefont {Kangkang}\ \bibnamefont {{Li}}}, \bibinfo {author}
  {\bibfnamefont {Chen}\ \bibnamefont {{Fang}}}, \ and\ \bibinfo {author}
  {\bibfnamefont {Yuan}\ \bibnamefont {{Li}}},\ }\bibfield  {title} {\enquote
  {\bibinfo {title} {{Topological spin excitations in a three-dimensional
  antiferromagnet}},}\ }\href {\doibase 10.1038/s41567-018-0213-x} {\bibfield
  {journal} {\bibinfo  {journal} {Nature Physics}\ }\textbf {\bibinfo {volume}
  {14}},\ \bibinfo {pages} {1011--1015} (\bibinfo {year} {2018})},\ \Eprint
  {http://arxiv.org/abs/1711.00632} {arXiv:1711.00632 [cond-mat.mes-hall]}
  \BibitemShut {NoStop}%
\bibitem [{\citenamefont {{Lu}}\ and\ \citenamefont {{Lu}}(2018)}]{Lu2018}%
  \BibitemOpen
  \bibfield  {author} {\bibinfo {author} {\bibfnamefont {Fuyan}\ \bibnamefont
  {{Lu}}}\ and\ \bibinfo {author} {\bibfnamefont {Yuan-Ming}\ \bibnamefont
  {{Lu}}},\ }\bibfield  {title} {\enquote {\bibinfo {title} {{Magnon band
  topology in spin-orbital coupled magnets: classification and application to
  $\alpha$-RuCl$_3$}},}\ }\href@noop {} {\bibfield  {journal} {\bibinfo
  {journal} {arXiv e-prints}\ ,\ \bibinfo {eid} {arXiv:1807.05232}} (\bibinfo
  {year} {2018})},\ \Eprint {http://arxiv.org/abs/1807.05232} {arXiv:1807.05232
  [cond-mat.str-el]} \BibitemShut {NoStop}%
\bibitem [{\citenamefont {Bao}\ \emph {et~al.}(2018)\citenamefont {Bao},
  \citenamefont {Wang}, \citenamefont {Wang}, \citenamefont {Cai},
  \citenamefont {Li}, \citenamefont {Ma}, \citenamefont {Wang}, \citenamefont
  {Ran}, \citenamefont {Dong}, \citenamefont {Abernathy}, \citenamefont {Yu},
  \citenamefont {Wan}, \citenamefont {Li},\ and\ \citenamefont
  {Wen}}]{Bao2018}%
  \BibitemOpen
  \bibfield  {author} {\bibinfo {author} {\bibfnamefont {Song}\ \bibnamefont
  {Bao}}, \bibinfo {author} {\bibfnamefont {Jinghui}\ \bibnamefont {Wang}},
  \bibinfo {author} {\bibfnamefont {Wei}\ \bibnamefont {Wang}}, \bibinfo
  {author} {\bibfnamefont {Zhengwei}\ \bibnamefont {Cai}}, \bibinfo {author}
  {\bibfnamefont {Shichao}\ \bibnamefont {Li}}, \bibinfo {author}
  {\bibfnamefont {Zhen}\ \bibnamefont {Ma}}, \bibinfo {author} {\bibfnamefont
  {Di}~\bibnamefont {Wang}}, \bibinfo {author} {\bibfnamefont {Kejing}\
  \bibnamefont {Ran}}, \bibinfo {author} {\bibfnamefont {Zhao-Yang}\
  \bibnamefont {Dong}}, \bibinfo {author} {\bibfnamefont {D.~L.}\ \bibnamefont
  {Abernathy}}, \bibinfo {author} {\bibfnamefont {Shun-Li}\ \bibnamefont {Yu}},
  \bibinfo {author} {\bibfnamefont {Xiangang}\ \bibnamefont {Wan}}, \bibinfo
  {author} {\bibfnamefont {Jian-Xin}\ \bibnamefont {Li}}, \ and\ \bibinfo
  {author} {\bibfnamefont {Jinsheng}\ \bibnamefont {Wen}},\ }\bibfield  {title}
  {\enquote {\bibinfo {title} {Discovery of coexisting dirac and triply
  degenerate magnons in a three-dimensional antiferromagnet},}\ }\href
  {\doibase 10.1038/s41467-018-05054-2} {\bibfield  {journal} {\bibinfo
  {journal} {Nature Communications}\ }\textbf {\bibinfo {volume} {9}},\
  \bibinfo {pages} {2591} (\bibinfo {year} {2018})}\BibitemShut {NoStop}%
\bibitem [{\citenamefont {Varga}\ \emph {et~al.}(2012)\citenamefont {Varga},
  \citenamefont {Droubay}, \citenamefont {Bowden}, \citenamefont {Nachimuthu},
  \citenamefont {Shutthanandan}, \citenamefont {Bolin}, \citenamefont
  {Shelton},\ and\ \citenamefont {Chambers}}]{Varga2012}%
  \BibitemOpen
  \bibfield  {author} {\bibinfo {author} {\bibfnamefont {Tamas}\ \bibnamefont
  {Varga}}, \bibinfo {author} {\bibfnamefont {Timothy~C.}\ \bibnamefont
  {Droubay}}, \bibinfo {author} {\bibfnamefont {Mark~E.}\ \bibnamefont
  {Bowden}}, \bibinfo {author} {\bibfnamefont {Ponnusamy}\ \bibnamefont
  {Nachimuthu}}, \bibinfo {author} {\bibfnamefont {Vaithiyalingam}\
  \bibnamefont {Shutthanandan}}, \bibinfo {author} {\bibfnamefont {Trudy~B.}\
  \bibnamefont {Bolin}}, \bibinfo {author} {\bibfnamefont {William~A.}\
  \bibnamefont {Shelton}}, \ and\ \bibinfo {author} {\bibfnamefont {Scott~A.}\
  \bibnamefont {Chambers}},\ }\bibfield  {title} {\enquote {\bibinfo {title}
  {Epitaxial growth of nitio3 with a distorted ilmenite structure},}\ }\href
  {\doibase https://doi.org/10.1016/j.tsf.2012.04.060} {\bibfield  {journal}
  {\bibinfo  {journal} {Thin Solid Films}\ }\textbf {\bibinfo {volume} {520}},\
  \bibinfo {pages} {5534 -- 5541} (\bibinfo {year} {2012})}\BibitemShut
  {NoStop}%
\bibitem [{\citenamefont {Newnham}\ \emph {et~al.}(1964)\citenamefont
  {Newnham}, \citenamefont {Fang},\ and\ \citenamefont
  {Santoro}}]{Newnham1964}%
  \BibitemOpen
  \bibfield  {author} {\bibinfo {author} {\bibfnamefont {R.~E.}\ \bibnamefont
  {Newnham}}, \bibinfo {author} {\bibfnamefont {J.~H.}\ \bibnamefont {Fang}}, \
  and\ \bibinfo {author} {\bibfnamefont {R.~P.}\ \bibnamefont {Santoro}},\
  }\bibfield  {title} {\enquote {\bibinfo {title} {{Crystal structure and
  magnetic properties of CoTiO${\sb 3}$}},}\ }\href {\doibase
  10.1107/S0365110X64000615} {\bibfield  {journal} {\bibinfo  {journal} {Acta
  Crystallographica}\ }\textbf {\bibinfo {volume} {17}},\ \bibinfo {pages}
  {240--242} (\bibinfo {year} {1964})}\BibitemShut {NoStop}%
\bibitem [{\citenamefont {Liu}\ and\ \citenamefont
  {Khaliullin}(2018)}]{Liu2018}%
  \BibitemOpen
  \bibfield  {author} {\bibinfo {author} {\bibfnamefont {Huimei}\ \bibnamefont
  {Liu}}\ and\ \bibinfo {author} {\bibfnamefont {Giniyat}\ \bibnamefont
  {Khaliullin}},\ }\bibfield  {title} {\enquote {\bibinfo {title} {Pseudospin
  exchange interactions in ${d}^{7}$ cobalt compounds: Possible realization of
  the kitaev model},}\ }\href {\doibase 10.1103/PhysRevB.97.014407} {\bibfield
  {journal} {\bibinfo  {journal} {Phys. Rev. B}\ }\textbf {\bibinfo {volume}
  {97}},\ \bibinfo {pages} {014407} (\bibinfo {year} {2018})}\BibitemShut
  {NoStop}%
\bibitem [{\citenamefont {Sano}\ \emph {et~al.}(2018)\citenamefont {Sano},
  \citenamefont {Kato},\ and\ \citenamefont {Motome}}]{Sano2018}%
  \BibitemOpen
  \bibfield  {author} {\bibinfo {author} {\bibfnamefont {Ryoya}\ \bibnamefont
  {Sano}}, \bibinfo {author} {\bibfnamefont {Yasuyuki}\ \bibnamefont {Kato}}, \
  and\ \bibinfo {author} {\bibfnamefont {Yukitoshi}\ \bibnamefont {Motome}},\
  }\bibfield  {title} {\enquote {\bibinfo {title} {Kitaev-heisenberg
  hamiltonian for high-spin ${d}^{7}$ mott insulators},}\ }\href {\doibase
  10.1103/PhysRevB.97.014408} {\bibfield  {journal} {\bibinfo  {journal} {Phys.
  Rev. B}\ }\textbf {\bibinfo {volume} {97}},\ \bibinfo {pages} {014408}
  (\bibinfo {year} {2018})}\BibitemShut {NoStop}%
\bibitem [{\citenamefont {Ross}\ \emph {et~al.}(2017)\citenamefont {Ross},
  \citenamefont {Brown}, \citenamefont {Cava}, \citenamefont {Krizan},
  \citenamefont {Nagler}, \citenamefont {Rodriguez-Rivera},\ and\ \citenamefont
  {Stone}}]{Ross2017}%
  \BibitemOpen
  \bibfield  {author} {\bibinfo {author} {\bibfnamefont {K.~A.}\ \bibnamefont
  {Ross}}, \bibinfo {author} {\bibfnamefont {J.~M.}\ \bibnamefont {Brown}},
  \bibinfo {author} {\bibfnamefont {R.~J.}\ \bibnamefont {Cava}}, \bibinfo
  {author} {\bibfnamefont {J.~W.}\ \bibnamefont {Krizan}}, \bibinfo {author}
  {\bibfnamefont {S.~E.}\ \bibnamefont {Nagler}}, \bibinfo {author}
  {\bibfnamefont {J.~A.}\ \bibnamefont {Rodriguez-Rivera}}, \ and\ \bibinfo
  {author} {\bibfnamefont {M.~B.}\ \bibnamefont {Stone}},\ }\bibfield  {title}
  {\enquote {\bibinfo {title} {Single-ion properties of the
  ${S}_{\mathrm{eff}}$ = $\frac{1}{2}$ xy antiferromagnetic pyrochlores
  $\mathrm{Na}{A}^{\ensuremath{'}}{\mathrm{co}}_{2}{\mathrm{f}}_{7}$
  (${A}^{\ensuremath{'}}={\mathrm{ca}}^{2+}, {\mathrm{sr}}^{2+}$)},}\ }\href
  {\doibase 10.1103/PhysRevB.95.144414} {\bibfield  {journal} {\bibinfo
  {journal} {Phys. Rev. B}\ }\textbf {\bibinfo {volume} {95}},\ \bibinfo
  {pages} {144414} (\bibinfo {year} {2017})}\BibitemShut {NoStop}%
\bibitem [{\citenamefont {Watanabe}\ \emph {et~al.}(1980)\citenamefont
  {Watanabe}, \citenamefont {Yamauchi},\ and\ \citenamefont
  {Takei}}]{Wanatabe1980}%
  \BibitemOpen
  \bibfield  {author} {\bibinfo {author} {\bibfnamefont {H.}~\bibnamefont
  {Watanabe}}, \bibinfo {author} {\bibfnamefont {H.}~\bibnamefont {Yamauchi}},
  \ and\ \bibinfo {author} {\bibfnamefont {H.}~\bibnamefont {Takei}},\
  }\bibfield  {title} {\enquote {\bibinfo {title} {Magnetic anisotropies in
  mtio3 (m = co, ni)},}\ }\href {\doibase
  https://doi.org/10.1016/0304-8853(80)90658-7} {\bibfield  {journal} {\bibinfo
   {journal} {Journal of Magnetism and Magnetic Materials}\ }\textbf {\bibinfo
  {volume} {15-18}},\ \bibinfo {pages} {549 -- 550} (\bibinfo {year}
  {1980})}\BibitemShut {NoStop}%
\bibitem [{\citenamefont {Kato}\ \emph {et~al.}(1986)\citenamefont {Kato},
  \citenamefont {Yamaguchi}, \citenamefont {Yamada}, \citenamefont {Funahashi},
  \citenamefont {Nakagawa},\ and\ \citenamefont {Takei}}]{Kato1986}%
  \BibitemOpen
  \bibfield  {author} {\bibinfo {author} {\bibfnamefont {H}~\bibnamefont
  {Kato}}, \bibinfo {author} {\bibfnamefont {Y}~\bibnamefont {Yamaguchi}},
  \bibinfo {author} {\bibfnamefont {M}~\bibnamefont {Yamada}}, \bibinfo
  {author} {\bibfnamefont {S}~\bibnamefont {Funahashi}}, \bibinfo {author}
  {\bibfnamefont {Y}~\bibnamefont {Nakagawa}}, \ and\ \bibinfo {author}
  {\bibfnamefont {H}~\bibnamefont {Takei}},\ }\bibfield  {title} {\enquote
  {\bibinfo {title} {Neutron scattering study of magnetic excitations in
  oblique easy-axis antiferromagnet {FeTiO}3},}\ }\href {\doibase
  10.1088/0022-3719/19/35/013} {\bibfield  {journal} {\bibinfo  {journal}
  {Journal of Physics C: Solid State Physics}\ }\textbf {\bibinfo {volume}
  {19}},\ \bibinfo {pages} {6993--7011} (\bibinfo {year} {1986})}\BibitemShut
  {NoStop}%
\bibitem [{\citenamefont {Goodenough}\ and\ \citenamefont
  {Stickler}(1967)}]{Goodenough1967}%
  \BibitemOpen
  \bibfield  {author} {\bibinfo {author} {\bibfnamefont {John~B.}\ \bibnamefont
  {Goodenough}}\ and\ \bibinfo {author} {\bibfnamefont {John~J.}\ \bibnamefont
  {Stickler}},\ }\bibfield  {title} {\enquote {\bibinfo {title} {Theory of the
  magnetic properties of the ilmenites $m\mathrm{Ti}{\mathrm{o}}_{3}$},}\
  }\href {\doibase 10.1103/PhysRev.164.768} {\bibfield  {journal} {\bibinfo
  {journal} {Phys. Rev.}\ }\textbf {\bibinfo {volume} {164}},\ \bibinfo {pages}
  {768--778} (\bibinfo {year} {1967})}\BibitemShut {NoStop}%
\bibitem [{\citenamefont {Balbashov}\ \emph {et~al.}(2017)\citenamefont
  {Balbashov}, \citenamefont {Mukhin}, \citenamefont {Ivanov}, \citenamefont
  {Iskhakova},\ and\ \citenamefont {Voronchikhina}}]{balbashov2017}%
  \BibitemOpen
  \bibfield  {author} {\bibinfo {author} {\bibfnamefont {A.~M.}\ \bibnamefont
  {Balbashov}}, \bibinfo {author} {\bibfnamefont {A.~A.}\ \bibnamefont
  {Mukhin}}, \bibinfo {author} {\bibfnamefont {V.~Yu.}\ \bibnamefont {Ivanov}},
  \bibinfo {author} {\bibfnamefont {L.~D.}\ \bibnamefont {Iskhakova}}, \ and\
  \bibinfo {author} {\bibfnamefont {M.~E.}\ \bibnamefont {Voronchikhina}},\
  }\bibfield  {title} {\enquote {\bibinfo {title} {Electric and magnetic
  properties of titanium-cobalt-oxide single crystals produced by floating zone
  melting with light heating},}\ }\href {\doibase 10.1063/1.5001297} {\bibfield
   {journal} {\bibinfo  {journal} {Low Temperature Physics}\ }\textbf {\bibinfo
  {volume} {43}},\ \bibinfo {pages} {965--970} (\bibinfo {year} {2017})},\
  \Eprint {http://arxiv.org/abs/https://doi.org/10.1063/1.5001297}
  {https://doi.org/10.1063/1.5001297} \BibitemShut {NoStop}%
\bibitem [{Kha()}]{Khait2019}%
  \BibitemOpen
  \href@noop {} {}\bibinfo {note} {I. Khait, G. Massarelli, S. Sorn, B. Yuan,
  Y.-J. Kim, A. Paramekanti, manuscript in preparation}\BibitemShut {NoStop}%
\bibitem [{\citenamefont {Guinea}\ \emph {et~al.}(2009)\citenamefont {Guinea},
  \citenamefont {Katsnelson},\ and\ \citenamefont {Geim}}]{Guinea2009}%
  \BibitemOpen
  \bibfield  {author} {\bibinfo {author} {\bibfnamefont {F.}~\bibnamefont
  {Guinea}}, \bibinfo {author} {\bibfnamefont {M.~I.}\ \bibnamefont
  {Katsnelson}}, \ and\ \bibinfo {author} {\bibfnamefont {A.~K.}\ \bibnamefont
  {Geim}},\ }\bibfield  {title} {\enquote {\bibinfo {title} {Energy gaps and a
  zero-field quantum hall effect in graphene by strain engineering},}\ }\href
  {\doibase 10.1038/nphys1420} {\bibfield  {journal} {\bibinfo  {journal}
  {Nature Physics}\ }\textbf {\bibinfo {volume} {6}},\ \bibinfo {pages} {30}
  (\bibinfo {year} {2009})}\BibitemShut {NoStop}%
\bibitem [{\citenamefont {Yelon}\ and\ \citenamefont
  {Silberglitt}(1971)}]{Yelon1971}%
  \BibitemOpen
  \bibfield  {author} {\bibinfo {author} {\bibfnamefont {W.~B.}\ \bibnamefont
  {Yelon}}\ and\ \bibinfo {author} {\bibfnamefont {Richard}\ \bibnamefont
  {Silberglitt}},\ }\bibfield  {title} {\enquote {\bibinfo {title}
  {Renormalization of large-wave-vector magnons in ferromagnetic
  cr${\mathrm{br}}_{3}$ studied by inelastic neutron scattering: Spin-wave
  correlation effects},}\ }\href {\doibase 10.1103/PhysRevB.4.2280} {\bibfield
  {journal} {\bibinfo  {journal} {Phys. Rev. B}\ }\textbf {\bibinfo {volume}
  {4}},\ \bibinfo {pages} {2280--2286} (\bibinfo {year} {1971})}\BibitemShut
  {NoStop}%
\bibitem [{\citenamefont {{Cao}}\ \emph {et~al.}(2018)\citenamefont {{Cao}},
  \citenamefont {{Fatemi}}, \citenamefont {{Fang}}, \citenamefont {{Watanabe}},
  \citenamefont {{Taniguchi}}, \citenamefont {{Kaxiras}},\ and\ \citenamefont
  {{Jarillo Herrero}}}]{Cao2018}%
  \BibitemOpen
  \bibfield  {author} {\bibinfo {author} {\bibfnamefont {Yuan}\ \bibnamefont
  {{Cao}}}, \bibinfo {author} {\bibfnamefont {Valla}\ \bibnamefont {{Fatemi}}},
  \bibinfo {author} {\bibfnamefont {Shiang}\ \bibnamefont {{Fang}}}, \bibinfo
  {author} {\bibfnamefont {Kenji}\ \bibnamefont {{Watanabe}}}, \bibinfo
  {author} {\bibfnamefont {Takashi}\ \bibnamefont {{Taniguchi}}}, \bibinfo
  {author} {\bibfnamefont {Efthimios}\ \bibnamefont {{Kaxiras}}}, \ and\
  \bibinfo {author} {\bibfnamefont {Pablo}\ \bibnamefont {{Jarillo Herrero}}},\
  }\bibfield  {title} {\enquote {\bibinfo {title} {{Unconventional
  superconductivity in magic-angle graphene superlattices}},}\ }\href {\doibase
  10.1038/nature26160} {\bibfield  {journal} {\bibinfo  {journal} {Nature}\
  }\textbf {\bibinfo {volume} {556}},\ \bibinfo {pages} {43--50} (\bibinfo
  {year} {2018})}\BibitemShut {NoStop}%
\end{thebibliography}

%

\end{document}